\documentclass{ws-procs10x7}

\usepackage{graphicx}
\usepackage[nosort]{cite}

\catcode`\@=11
\def\@cite#1#2{\unskip\nobreak\relax
    \def\@tempa{$\m@th^{\hbox{\the\scriptfont0 #1}}$}%
    \futurelet\@tempc\@citexx}
\def\@citexx{\ifx.\@tempc\let\@tempd=\@citepunct\else
    \ifx,\@tempc\let\@tempd=\@citepunct\else
    \ifx;\@tempc\let\@tempd=\@citepunct\else
    \let\@tempd=\@tempa\fi\fi\fi\@tempd}
\def\@citepunct{\@tempc\edef\@sf{\spacefactor=\the\spacefactor\relax}\@tempa
    \@sf\@gobble}
\catcode`@=12

\def\OMIT#1{{}}

\def\vereq#1#2{\lower3.5pt\vbox{\baselineskip1.5pt \lineskip1.5pt
\ialign{$#1\hfill##\hfil$\crcr#2\crcr\sim\crcr}}}

\def\lqcd{\Lambda_{\rm QCD}}

\def\ov{\overline}

\def\fbmo{\ensuremath{{\rm fb}^{-1}}}

\newcommand{\Bbar}{\,\overline{\!B}}
\newcommand{\Dbar}{\,\overline{\!D}}
\newcommand{\Kbar}{\,\overline{\!K}}
\def\B0bar{\Bbar{}^0}
\def\D0bar{\Dbar{}^0}
\def\K0bar{\Kbar{}^0}

\newcommand{\nn}{\nonumber}
\newcommand{\beq}{\begin{equation}}
\newcommand{\eeq}{\end{equation}}
\newcommand{\beqa}{\begin{eqnarray}}
\newcommand{\eeqa}{\end{eqnarray}}

\begin{document}

\title{\boldmath The CKM matrix and $CP$ Violation}

\author{Zoltan Ligeti}

\address{Ernest Orlando Lawrence Berkeley National Laboratory\\
  University of California, Berkeley, CA 94720\\ 
  E-mail: zligeti@lbl.gov}

\twocolumn[\maketitle\abstract{
The status of $CP$ violation and the CKM matrix is reviewed.  Direct $CP$
violation in $B$ decay has been established and the measurement of $\sin2\beta$
in $\psi K$ modes reached $5\%$ accuracy.  I discuss the implications of these,
and of the possible deviations of the $CP$ asymmetries in $b\to s$ modes from
that in $\psi K$.  The first meaningful measurements of $\alpha$ and $\gamma$
are explained, together with their significance for constraining both the SM and
new physics in $B-\Bbar$ mixing.  I also discuss implications of recent
developments in the theory of nonleptonic decays for $B\to \pi K$ rates and $CP$
asymmetries, and for the polarization in charmless $B$ decays to two vector
mesons.
\hfill {\footnotesize LBNL-55944}}]

\section{Introduction}

In the last few years the study of $CP$ violation and flavor physics has
undergone dramatic developments.  While for 35 years, until 1999, the only
unambiguous measurement of $CP$ violation (CPV) was $\epsilon_K$~\cite{Kcpv},
the constraints on the CKM matrix~\cite{C,KM} improved tremendously since the
$B$ factories turned on.  The error of $\sin2\beta$ is an order of magnitude
smaller now than in the first measurements few years ago [see Eq.~(\ref{s2b})].

Flavor and $CP$ violation are excellent probes of new physics (NP), as
demonstrated by the following examples:
\begin{itemize}\vspace*{-6pt}\itemsep -2pt
\item Absence of $K_L\to \mu\mu$ predicted charm;
\item $\epsilon_K$ predicted the third generation;
\item $\Delta m_K$ predicted the charm mass;
\item $\Delta m_B$ predicted the heavy top mass.
\end{itemize}\vspace*{-6pt}
From these measurements we know already that if there is NP at the TeV scale
then it must have a very special flavor and $CP$ structure to satisfy the
existing constraints.

The question we would like to address is: What does the new data tell us?

\subsection{Testing the flavor sector}

In the SM only the Yukawa couplings distinguish between the fermion
generations.   This is a coupling to something unknown, which we would like to
understand better.  In the SM there are 10 physical quark flavor parameters, the
6 quark masses and the 4 parameters in the CKM matrix: 3 mixing angles and 1
$CP$ violating phase~\cite{Ligeti:2003fi}.  Therefore, the SM predicts intricate
correlations between dozens of different decays of $s$, $c$, $b$, and $t$
quarks, and in particular between $CP$ violating observables.  Possible
deviations from CKM paradigm may upset some predictions:
\begin{itemize} \vspace*{-6pt}\itemsep -2pt
\item Flavor-changing neutral currents at unexpected level, e.g., $B_s$ mixing
incompatible with SM,  enhanced $B_{(s)}\to \ell^+ \ell^-$;
\item Subtle (or not so subtle) changes in correlations, e.g., $CP$ asymmetries
not equal in $B\to \psi K_S$ and $B\to \phi K_S$;
\item Enhanced or suppressed $CP$ violation, e.g., $B_s\to \psi \phi$.
\end{itemize} \vspace*{-4pt}

\begin{figure}[t]
\centerline{\includegraphics*[width=.4\textwidth]{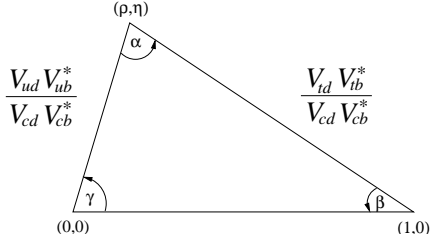}}
\caption{Sketch of the unitarity triangle.}
\label{fig:triangle}
\end{figure}

The key to testing the SM is to do many overconstraining measurements. A
convenient language to compare these is by putting constraints on $\rho$ and
$\eta$, which occur in the Wolfenstein parameterization of the CKM matrix,
\beqa\label{ckmdef}
V_{\rm CKM} = \pmatrix{ V_{ud} & V_{us} & V_{ub} \cr
  V_{cd} & V_{cs} & V_{cb} \cr
  V_{td} & V_{ts} & V_{tb} } = \hspace*{1.7cm} \\[2pt]
\pmatrix{ 1-\frac{1}{2}\lambda^2 & \lambda & A\lambda^3(\rho-i\eta) \cr
  -\lambda & \!\!1-\frac{1}{2}\lambda^2\!\! & A\lambda^2 \cr
  A\lambda^3(1-\rho-i\eta) & -A\lambda^2 & 1} . \hspace*{-.25cm} \nn
\eeqa
This form is designed to exhibit the hierarchical structure by expanding in the
sine of the Cabibbo angle, $\lambda = \sin\theta_C \simeq 0.22$, and is valid to
order $\lambda^4$.  The unitarity of $V_{\rm CKM}$ implies several relations,
such as
\beq
V_{ud}\, V_{ub}^* + V_{cd}\, V_{cb}^* + V_{td}\, V_{tb}^* = 0 \,.
\eeq
A graphical representation of this is the unitarity triangle, obtained by
rescaling the best-known side to unit length (see Fig.~\ref{fig:triangle}). Its
sides and angles can be determined in many ``redundant" ways, by measuring $CP$
violating and conserving observables.

\subsection{Constraints from $K$ and $D$ decays}
\label{sec:KD}

We know from the measurement of $\epsilon_K$ that CPV in the $K$ system is at
the right level, as it can be accommodated in the SM with an ${\cal O}(1)$
value of the KM phase~\cite{KM}.  The other observed $CP$ violating quantity in
kaon decay, $\epsilon_K'$, is notoriously hard to calculate, so hadronic
uncertainties have precluded precision tests of the KM mechanism.  In the kaon
sector these will come from the study of $K\to \pi\nu\bar\nu$ decays.  The BNL
E949 experiment observed the third event, yielding~\cite{Anisimovsky:2004hr}
\beq
{\cal B}(K^+\to \pi^+\nu\bar\nu) = (1.47^{+1.30}_{-0.89}) \times 10^{-10}\,.
\eeq
This is consistent with the SM within the large uncertainties, but much more
statistics is needed to make definitive tests.

The $D$ meson system is complementary to $K$ and $B$ mesons, because flavor and
$CP$ violation are suppressed both by the GIM mechanism and by the Cabibbo
angle.  Therefore, CPV in $D$ decays, rare $D$ decays, and $D-\Dbar$ mixing are
predicted to be small in the SM and have not been observed.  The $D^0-\D0bar$ is
the only neutral meson mixing generated by down-type quarks in the SM (up-type
squarks in SUSY).  The strongest hint for $D^0-\D0bar$ mixing
is~\cite{Yabsley:2003rn}
\beqa
y_{CP} &=& {\Gamma(CP \mbox{ even}) - \Gamma(CP \mbox{ odd}) \over
  \Gamma(CP \mbox{ even}) + \Gamma(CP \mbox{ odd})} \nn\\ 
&=& (0.9 \pm 0.4)\%\,.
\eeqa
Unfortunately, because of hadronic uncertainties, this measurement cannot be
interpreted as a sign of new physics~\cite{Falk:2001hx}.  At the present level
of sensitivity, CPV would be the only clean signal of NP in the $D$ sector.

\section{\boldmath $CP$ violation in $B$ decays and $B\to J/\psi K_S$}
\label{sec:Bbc}

\subsection{$CP$ violation in decay}
\label{CPVdecay}

$CP$ violation in decay is in some sense its simplest form, and can 
be observed in both charged and neutral meson as well as in baryon decays.  It
requires at least two amplitudes with nonzero relative weak $(\phi_k)$ and
strong $(\delta_k)$ phases to contribute to a decay,
\beqa
A_f &=& \langle f | {\cal H} |B\rangle = 
  \sum_k A_k\, e^{i\delta_k}\, e^{i\phi_k}\,, \nn\\
\ov{A}_{\ov f} &=& \langle \ov f | {\cal H} |\Bbar\rangle = 
  \sum_k A_k\, e^{i\delta_k}\, e^{-i\phi_k}\,.
\eeqa
If $|{\ov A_{\ov f} / A_f}| \neq 1$ then $CP$ is violated.

This type of $CP$ violation is unambiguously observed in the kaon sector by
$\epsilon'_K \neq 0$, and now it is also established in $B$ decays with
$5.7\sigma$ significance,
\beqa\label{KpidirectCP}
A_{K^-\pi^+} &\equiv& {\Gamma(\Bbar\to K^-\pi^+) - \Gamma(B\to K^+\pi^-)
  \over \Gamma(\Bbar\to K^-\pi^+) + \Gamma(B\to K^+\pi^-)} \nn\\
&=& -0.109 \pm 0.019\,,
\eeqa
averaging the BABAR~\cite{Aubert:2004qm}, BELLE~\cite{bellehh},
CDF~\cite{CDFKpi}, and CLEO~\cite{Chen:2000hv} measurements.  This is simply a
counting experiment: there are a significantly larger number of $B^0\to
K^+\pi^-$ than $\B0bar\to K^-\pi^+$ decays.

This measurement implies that after the ``$K$-superweak" model~\cite{superweak},
now also ``$B$-superweak" models are excluded.  I.e., models in which $CP$
violation in the $B$ sector only occurs in $B^0 - \B0bar$ mixing are no longer
viable.  This measurement also establishes that there are sizable strong phases
between the tree $(T)$ and penguin $(P)$ amplitudes in charmless $B$ decays,
since estimates of $|T/P|$ are not much larger than $A_{K^-\pi^+}$.  (Note that
a sizable strong phase has also been established in $B\to \psi
K^*$~\cite{Verderi:2004jp,higuchi}.)  Such information on strong phases will
have broader implications for the theory of charmless nonleptonic decays and for
understanding the $B\to K\pi$ and $\pi\pi$ rates discussed in
Sec.~\ref{sec:Kpi}.  

The bottom line is that, similar to $\epsilon'_K$, our theoretical understanding
at present is insufficient to either prove or rule out that the $CP$ asymmetry
in Eq.~(\ref{KpidirectCP}) is due to NP.

\subsection{CPV in mixing}

The two $B$ meson mass eigenstates are related to the flavor eigenstates via
\beq
|B_{L,H}\rangle = p |B^0\rangle \pm q |\B0bar\rangle\,.
\eeq
$CP$ is violated if the mass eigenstates are not equal to the $CP$ eigenstates.
This happens if $|q/p| \neq 1$, i.e., if the physical states are not
orthogonal, $\langle B_H | B_L\rangle \neq 0$, showing that this is an
intrinsically quantum mechanical phenomenon.

The simplest example of this type of $CP$ violation is the semileptonic decay
asymmetry to ``wrong sign" leptons,
\beqa\label{ASL}
A_{\rm SL} &=& {\Gamma(\B0bar(t) \to \ell^+ X) - \Gamma(B^0(t) \to \ell^- X)
  \over \Gamma(\B0bar(t) \to \ell^+ X) + \Gamma(B^0(t) \to \ell^- X) } \nn\\
&=& {1 - |q/p|^4 \over 1 + |q/p|^4} 
  = (-0.05 \pm 0.71)\% \,,
\eeqa
implying $|q/p| = 1.0003 \pm 0.0035$, where this average is dominated by a new
BELLE result~\cite{Abe:2004wh}.  In kaon decays the similar asymmetry has been
measured~\cite{cplear}, in agreement with the expectation that it is equal to
$4\,{\rm Re}\, \epsilon$.

The calculation of $A_{\rm SL}$ is only possible from first principles in the
$m_b \gg \lqcd$ limit using an operator product expansion to evaluate the
relevant nonleptonic rates.  The calculation has sizable uncertainties by virtue
of our limited understanding of $b$ hadron lifetimes.  Last year the NLO QCD
calculation of $A_{\rm SL}$ was completed~\cite{Beneke:2003az,Ciuchini:2003ww},
predicting $A_{\rm SL} = -(5.5 \pm 1.3)\times 10^{-4}$, where I averaged the
central values and quoted the larger of the two theory error estimates.  (The
similar asymmetry in the $B_s$ sector is expected to be $\lambda^2$ smaller.) 
Although the experimental error in Eq.~(\ref{ASL}) is an order of magnitude
larger than the SM expectation, this measurement already constraints new
physics~\cite{Laplace:2002ik}, as the $m_c^2/m_b^2$ suppression of $A_{\rm SL}$
in the SM can be avoided by NP.

\subsection{CPV in the interference between decay with and without mixing, $B\to
J/\psi K_S$ and its implications}

It is possible to obtain theoretically clean information on weak phases in  $B$
decays to certain $CP$ eigenstate final states.  The  interference phenomena
between $B^0\to f_{CP}$ and $B^0 \to \B0bar \to f_{CP}$ is described by
\beq
\lambda_{f_{CP}} = \frac qp\, \frac{\ov A_{f_{CP}}}{A_{f_{CP}}} 
  = \eta_{f_{CP}}\, \frac qp\, \frac{\ov{A}_{f_{CP}}}{A_{\ov{f}_{CP}}} \,,
\eeq
where $\eta_{f_{CP}} = \pm 1$ is the $CP$ eigenvalue of $f_{CP}$. 
Experimentally one can study the time dependent $CP$ asymmetry,
\beqa
a_{f_{CP}} &=& {\Gamma[\B0bar(t)\to f] - \Gamma[B^0(t)\to f]\over
  \Gamma[\B0bar(t)\to f] + \Gamma[B^0(t)\to f] }\qquad \\
&=& S_{f_{CP}} \sin(\Delta m\, t) - C_{f_{CP}} \cos(\Delta m\, t)\,,\nn
\eeqa
where
\beq
S_f = {2\,{\rm Im}\,\lambda_f\over
  1+|\lambda_f|^2}\,, \quad
C_f (= - A_f) = {1-|\lambda_f|^2 \over 1+|\lambda_f|^2}\,.
\eeq
If amplitudes with one weak phase dominate a decay then $a_{f_{CP}}$ measures a
phase in the Lagrangian theoretically cleanly.  In this case $C_f=0$, and 
$a_{f_{CP}} = {\rm Im}\,\lambda_f\, \sin(\Delta m\, t)$, where ${\rm
arg}\lambda_f$ is the phase difference between the two decay paths (with or
without mixing).

The theoretically cleanest example of this type of $CP$ violation is $B\to \psi
K_S$.  While there are tree and penguin contributions to the decay with
different weak phases, the dominant part of the penguin amplitudes have the same
weak phase as the tree amplitude.  Therefore, contributions with the tree
amplitude's weak phase dominate, to an accuracy better than $\sim$1\%.  In the
usual phase convention ${\rm arg}\,\lambda_{\psi K_S} = (B\mbox{-mixing}=2\beta)
+ (\mbox{decay}=0)  + (K\mbox{-mixing}=0)$, so we expect $S_{\psi K} =
\sin2\beta$ and $C_{\psi K} = 0$ to a similar accuracy.  The new world average
is
\beq\label{s2b}
\sin2\beta = 0.726 \pm 0.037\,,
\eeq
which is now a 5\% measurement.  For the first time $\cos 2\beta$ has also been
constrained, by studying angular distributions in the time dependent $B\to \psi
K^{*0}$ analysis.  BABAR obtained~\cite{Verderi:2004jp} $\cos 2\beta = +2.72
^{+0.50}_{-0.79} \pm 0.27$, excluding the negative $\cos 2\beta$ solution at the
89\% CL.  With more data, this will eliminate 2 of the 4 discrete ambiguities,
corresponding to $\beta = (\pi - \arcsin S_{\psi K})/2$ and $\beta = (3\pi -
\arcsin S_{\psi K})/2$.

\begin{figure}[t]
\centerline{\includegraphics[width=0.45\textwidth]{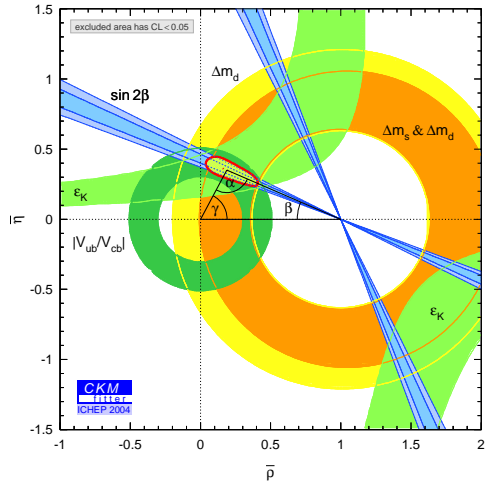}}
\caption{The present CKM fit.}
\label{fig:smplot}
\end{figure}

$S_{\psi K}$ was the first observation of $CP$ violation outside the kaon
sector, and the first observation of an ${\cal O}(1)$ effect that violates
$CP$.  It implies that models with approximate $CP$ symmetry (in the sense that
all CPV phases are small) are excluded.  The constraints on the CKM matrix from
the measurements of $S_{\psi K}$, $|V_{ub}/V_{cb}|$, $\epsilon_K$, $B$ and $B_s$
mixing are shown in Fig.~\ref{fig:smplot} using the CKMfitter
package~\cite{ckmfitter,Charles:2004jd}.  The overall consistency between these
measurements constitutes the first precise test of the CKM picture.  It also
implies that it is unlikely that we will find ${\cal O}(1)$ deviations from the
SM, and we should look for corrections rather than alternatives of the CKM
picture.

\section{\boldmath Other $CP$ asymmetries that are approximately
$\sin2\beta$ in the SM}
\label{sec:Bbs}

\begin{table*}[t]
\caption{$CP$ asymmetries for which the SM predicts $-\eta_f S_f \approx
\sin2\beta$.  The 3rd column contains my estimates of limits on the deviations
from $\sin2\beta$ in the SM (strict bounds are worse), and the last two columns
show the world averages~\cite{hfag}. (The $CP$-even fractions in $K^+ K^- K_S$
and $D^{*+}D^{*-}$ are determined experimentally.)}
\label{tab:exp_fCP}
\centerline{\begin{tabular}{|ccccc|}  
\hline
Dominant  &  final  &  SM upper limit on
  &  \raisebox{-6pt}[0pt][-6pt]{$-\eta_{f_{CP}} S_{f_{CP}}$}
  &  \raisebox{-6pt}[0pt][-6pt]{$C_{f_{CP}}$} 
\\[-2pt]
process  &  state  &  $|-\eta_{f_{CP}} S_{f_{CP}} - \sin2\beta|$  &  & 
\\ \hline\hline
$b\to c \bar c s$  &  $\psi K_S$  &  $<0.01$  
  &  $+0.726 \pm 0.037$  &  $+0.031\pm0.029$
\\ \hline
$b\to c \bar c d$  &  $\psi \pi^0$  &  $\sim 0.2$
  &  $+0.40 \pm 0.33$  &  $+0.12\pm0.24$
\\ 
  &  $D^{*+}D^{*-}$  &  $\sim 0.2$
  &  $+0.20 \pm 0.32$  &  $+0.28\pm0.17$
\\ \hline
$b\to s \bar q q$  &  $\phi K^0$  &  $\sim 0.05$
  &  $+0.34 \pm 0.20$  &  $-0.04\pm0.17$
\\ 
  &  $\eta' K_S$  &  $\sim 0.1$
  &  $+0.41 \pm 0.11$  &  $-0.04\pm0.08$
\\ 
  &  $K^+ K^- K_S$  &  $\sim 0.15$
  &  $+0.53 \pm 0.17$  &  $+0.09\pm0.10$
\\   
  &  $\pi^0 K_S$  &  $\sim 0.15$
  &  $+0.34 \pm 0.28$  &  $+0.09\pm0.14$
\\
  &  $f_0 K_S$  &  $\sim 0.15$
  &  $+0.39 \pm 0.26$  &  $+0.14\pm0.22$
\\
  &  $\omega K_S$  &  $\sim 0.15$
  &  $+0.75 \pm 0.66$  &  $-0.26\pm0.50$
\\ \hline
\end{tabular}}
\end{table*}

The $b\to s$ transitions, such as $B\to \phi K_S$, $\eta' K_S$, $K^+ K^- K_S$,
etc., are dominated by one-loop (penguin) diagrams in the SM, and therefore new
physics could compete with the SM contributions~\cite{Grossman:1996ke}.  Using
CKM unitarity we can write the contributions to such decays as a term
proportional to $V_{cb} V_{cs}^*$ and another proportional to $V_{ub}
V_{us}^*$.  Since their ratio is ${\cal O}(\lambda^2) \sim 0.05$, we expect
amplitudes with one weak phase to dominate these decays as well.  Thus, in the
SM, the measurements of $-\eta_f S_f$ should agree with each other and with
$S_{\psi K}$ to an accuracy of order $\lambda^2 \sim 0.05$.  

If there is a SM and a NP contribution, the asymmetries depend on their relative
size and phase, which depend on hadronic matrix elements.  Since these are
mode-dependent, the asymmetries will, in general, be different between the
various modes, and different from $S_{\psi K}$.  One may also find $C_f$
substantially different from 0.  (NP would have to dominate over the SM
amplitude in order that the asymmetries become different from the SM and equal
to one other.)

The averages of the latest BABAR~\cite{hoecker} and BELLE~\cite{sakai} results
are shown in Table~\ref{tab:exp_fCP}.  The two data sets are more consistent
than before, so averaging them seems meaningful at this time.  The single
largest deviation from the SM is in the $\eta'K_S$ mode, 
\beq\label{etapdiff}
S_{\psi K} - S_{\eta'K_S} =  0.31 \pm 0.12\,,
\eeq
which is $2.6\sigma$.  The average $CP$ asymmetry in all $b\to s$ modes, which
also equals $S_{\psi K}$ in the SM, has a more significant deviation,
\beq\label{avediff}
S_{\psi K} - \langle -\eta_f S_{f(b\to s)} \rangle = 0.30\pm0.08\,.
\eeq
This $3.5\sigma$ effect comes from $2.7\sigma$ at BABAR and $2.4\sigma$ at
BELLE.  It is a less than $3.5\sigma$ signal for NP, because some of the modes
included may deviate significantly from $S_{\psi K}$ in the SM.  However, there
is another $3.1\sigma$ effect,
\beq\label{neat}
S_{\psi K} - \langle S_{\eta'K_S,\phi K_S} \rangle = 0.33 \pm 0.11\,.
\eeq

The entries in the third column in Table~\ref{tab:exp_fCP} show my estimates of
limits on the deviations from $S_{\psi K}$ in the SM.  The hadronic matrix
elements multiplying the generic ${\cal O}(0.05)$ suppression of the ``SM
pollution" are hard to bound model independently~\cite{Grossman:2003qp}, so
strict bounds are weaker, while model calculations tend to obtain smaller
limits.  I attempted to list reasonable benchmarks for each mode.

\subsection{Implications of the data}

To understand the significance of Eq.~(\ref{etapdiff}) and (\ref{neat}), note
that a conservative bound using $SU(3)$ flavor symmetry and (updated)
experimental limits on related modes gives~\cite{Grossman:2003qp,Chiang:2003rb}
$|S_{\psi K} - S_{\eta'K_S}| < 0.2$ in the SM.  Most other estimates obtain
bounds a factor of two smaller or even better (these are also more model
dependent).  Thus we can be confident that, if established at the $5\sigma$
level, $S_{\eta'K_S} \approx 0.4$ would be a sign of NP.  (The deviation of
$S_{\phi K_S}$ from $S_{\psi K}$ is now less than $2\sigma$, but there is room
for discovery, as the present central value of $S_{\phi K}$ with a smaller error
could still establish NP.)  The largest deviation from the SM at present is the
$3.5\sigma$ effect in $\langle -\eta_f S_{f(b\to s)} \rangle$.  Such a discovery
would exclude in addition to the SM, models with minimal flavor violation, and
universal SUSY models, such as gauge mediated SUSY breaking.

In the last few years the central value of $S_{\phi K_S}$ got closer to $S_{\psi
K}$, while $S_{\eta' K_S}$ got further from it, disfavoring models in which NP
enters $S_{\phi K_S}$  but not $S_{\eta' K_S}$.  This includes models of
parity-even NP, which would affect $B\to \phi K_S$ (odd $\to$ odd) but not $B\to
\eta' K_S$ (odd $\to$ even).  This happens, for example, in a
left-right-symmetric SUSY model, if the LRS breaking scale is high enough so
that direct effects from the $W_R$ sector are absent~\cite{Kagan:2004ia}.  This
scenario is disfavored also because the $K^+K^-K_S$ final state is $P$-odd, just
like $\phi K_S$.

\begin{figure}[t]
\centerline{\includegraphics[width=0.4\textwidth]{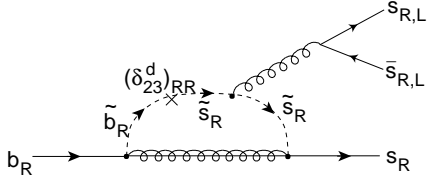}}
\caption{A SUSY contribution to $b\to s\bar s s$~\cite{Harnik:2002vs}.}
\label{fig:susypenguin}
\end{figure}

Model building may actually become more interesting with the new data.  The
present central values of $S_{\eta' K_S}$ and $S_{\phi K_S}$ can be reasonably
accommodated with NP, such as SUSY (unlike ${\cal O}(1)$ deviations from
$S_{\psi K_S}$).  While $B\to X_s\gamma$ mainly constrains $LR$ mass insertions,
penguins shown in Fig.~\ref{fig:susypenguin} involving $RR$ (and $LL$) mass
insertions can give sizable effect in $b\to s$ transitions.  However, as of this
conference, we also know that ${\cal B}(B\to X_s \ell^+\ell^-) = (4.5 \pm 1.0)
\times 10^{-6}$ agrees with the SM at the ${\cal O}(20\%)$ level~\cite{hfag},
which gives new constraints on the $RR$ and $LL$ mass insertions (replace $g \to
Z$ and $s\bar s \to \ell^+\ell^-$).

\section{\boldmath Measurements of $\alpha$ and $\gamma$}
\label{sec:alphagamma}

To clarify notation, I'll call a $\gamma$-measurement the determination of the
phase difference between $b\to u$ and $b\to c$ transitions, while $\alpha$ will
refer to the measurements of $\gamma$ in the presence of $B-\Bbar$ mixing
($\alpha \equiv \pi - \beta - \gamma$).  Interestingly, the methods that give
the best results were not even talked about before 2003.

\subsection{$\alpha$ from $B\to \pi\pi$}

In contrast to $B\to \psi K$, which is dominated by amplitudes with one week
phase, it is now well-established that in $B\to \pi^+\pi^-$ there are two
comparable contributions with different weak phases~\cite{Ali}.  Therefore, to
determine $\alpha$ model independently, it is necessary to carry out the 
isospin analysis~\cite{Gronau:1990ka}.  The hardest ingredient is the
measurement of the $\pi^0\pi^0$ mode,
\beq
{\cal B}(B\to \pi^0\pi^0) = (1.51 \pm 0.28) \times 10^{-6}\,,
\eeq
and in particular the need to measure the $CP$-tagged rates.
At this conference BABAR~\cite{cristinziani} and BELLE~\cite{bellehh} presented
the first such measurements, giving the world average
\beq\label{C00}
{\Gamma(\Bbar\to \pi^0\pi^0) - \Gamma(B\to \pi^0\pi^0)
  \over \Gamma(\Bbar\to \pi^0\pi^0) + \Gamma(B\to \pi^0\pi^0)}
  = 0.28 \pm 0.39.
\eeq

\begin{figure}[t]
\centerline{\includegraphics[width=.45\textwidth]{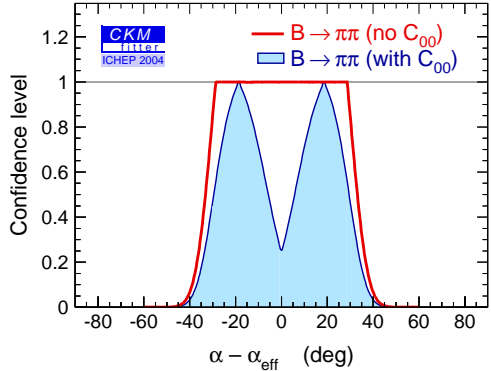}}
\caption{Constraints on $\alpha - \alpha_{\rm eff}$.}
\label{fig:apipi}
\end{figure}

Thus, for the first time, we can determine from the isospin analysis (with
sizable error) the penguin pollution, $\alpha - \alpha_{\rm eff}$ [$2\alpha_{\rm
eff} \equiv {\rm arg}\, \lambda_{\pi^+\pi^-} = \arcsin (S_{\pi^+\pi^-}
(1-C_{\pi^+\pi^-}^2)^{-1/2})$].  In Fig.~\ref{fig:apipi}, the blue (shaded)
region shows the confidence level using Eq.~(\ref{C00}), while the red (thick
solid) curve is the constraint without it.  We find $|\alpha - \alpha_{\rm eff}|
< 37^\circ$ at 90\% CL, a small improvement over the $39^\circ$ bound without
Eq.~(\ref{C00}); so it will take a lot more data to determine $\alpha$
precisely.  The interpretation for $\alpha$ is unclear at present, due to the
marginal consistency of the $S_{\pi^+\pi^-}$ data; see Table~\ref{tab:pipi}.

\begin{table}[h]
\caption{$CP$ violation in $B\to \pi^+\pi^-$.}
\label{tab:pipi}
\centerline{\begin{tabular}{|c|cc|}  
\hline
$B\to\pi^+\pi^-$  &  $S_{\pi^+\pi^-}$  &  $C_{\pi^+\pi^-}$ 
\\ \hline\hline
BABAR  &  $-0.30 \pm 0.17$  &  $-0.09 \pm 0.15$ \\
BELLE  &  $-1.00 \pm 0.22$  &  $-0.58 \pm 0.17$ \\
average  &  $-0.61 \pm 0.14$  &  $-0.37 \pm 0.11$ 
\\ \hline
\end{tabular}}
\end{table}

\subsection{$\alpha$ from $B\to \rho\rho$}

$B\to\rho\rho$ is more complicated than $B\to \pi\pi$ in that a vector-vector
($VV$) final state is a mixture of $CP$-even ($L=0$ and 2) and -odd ($L=1$)
components.  The $B\to \pi\pi$ isospin analysis applies for each $L$ in
$B\to\rho\rho$ (or in the transversity basis for each $\sigma = 0, \parallel,
\perp$).\break  The situation is simplified dramatically by the experimental
observation that in the $\rho^+\rho^-$ and $\rho^+\rho^0$ modes the longitudinal
polarization fraction is near unity (see Sec.~\ref{sec:pol}), so the $CP$-even
fraction dominates.  Thus, one can simply bound $\alpha - \alpha_{\rm eff}$
from~\cite{dallapiccola}
\beq
{\cal B}(B\to \rho^0\rho^0) < 1.1 \times 10^{-6}\ (90\%\ {\rm CL})\,.
\eeq
The smallness of this rate implies that $\alpha-\alpha_{\rm eff}$ in $B\to
\rho\rho$ is much smaller than in $B\to \pi\pi$.  To indicate the difference,
note that ${{\cal B}(B\to \pi^0\pi^0) / {\cal B}(B\to \pi^+\pi^0)} = 0.27 \pm
0.06$, while ${{\cal B}(B\to \rho^0\rho^0) / {\cal B}(B\to \rho^+\rho^0)} <
0.04\ \ (90\%\ {\rm CL})$.  From $S_{\rho^+\rho^-}$ and the isospin bound on
$\alpha-\alpha_{\rm eff}$ BABAR obtains~\cite{dallapiccola}
\beq
\alpha = 96 \pm 10 \pm 4 \pm 11^\circ(\alpha-\alpha_{\rm eff})\,.
\eeq
Ultimately the isospin analysis is more complicated in $B\to\rho\rho$ than in
$\pi\pi$, because the nonzero value of $\Gamma_\rho$ allows for the final state
to be in an isospin-1 state~\cite{Falk:2003uq}.  This only affects the results
at the ${\cal O}(\Gamma_\rho^2/m_\rho^2)$ level, which is smaller than other
errors at present.  With higher statistics, it will be possible to constrain
this effect using the data~\cite{Falk:2003uq}.

\subsection{$\alpha$ from $B\to \rho\pi$}

In the two-body analysis isospin symmetry gives two pentagon
relations~\cite{Lipkin:1991st}.  Solving them would require measurements of the
rates and $CP$ asymmetries in all the $B\to \rho^+\pi^-$,  $\rho^-\pi^+$, and
$\rho^0\pi^0$ modes, which is not available.  While BABAR set a 90\% CL
upper bound~\cite{Aubert:2003fm} ${\cal B}(B \to \rho^0\pi^0) < 2.9 \times
10^{-6}$, BELLE measured~\cite{Dragic:2004tj} ${\cal B}(B \to \rho^0\pi^0) =
(5.1 \pm 1.6 \pm 0.9) \times 10^{-6}$.  The two experiments agree on the direct
$CP$ asymmetries~\cite{cristinziani,Wang:2004va}, and their average
\beqa
A_{\pi^-\rho^+} &=& -0.48 ^{+0.13}_{-0.14}\,, \nn\\
A_{\pi^+\rho^-} &= &-0.15 \pm 0.09\,,
\eeqa
is $3.6\sigma$ from no direct $CP$ violation, $(A_{\pi^-\rho^+},$
$A_{\pi^+\rho^-}) = (0,0)$.  With assumptions about factorization and $SU(3)$
flavor symmetry, one can obtain $\alpha = (95 \pm 6_{\rm (exp)} \pm 15_{\rm
(th)})^\circ$~\cite{Gronau:2004tm}, but here the error is theory dominated.

At this conference BABAR showed the first Dalitz plot
analysis~\cite{Snyder:1993mx} of the the interference regions in $B\to
\pi^+\pi^-\pi^0$ to determine~\cite{cristinziani}
\beq
\alpha = (113 ^{+27}_{-17} \pm 6)^\circ ,
\eeq
which uses less assumptions than the extraction of $\alpha$ from the two-body
measurements.

\subsection{Combined determination of $\alpha$}

\begin{figure}[t]
\centerline{\includegraphics[width=.45\textwidth]{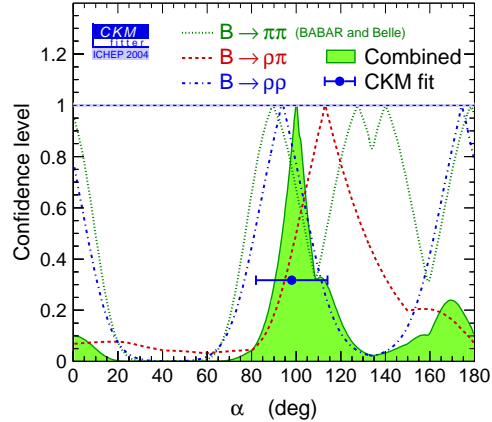}}
\caption{Confidence levels of the $\alpha$ measurements.}
\label{fig:alpha}
\end{figure}

The combination of these measurements of $\alpha$ is shown in
Fig.~\ref{fig:alpha}.  Due to the marginal consistency of the $S_{\pi^+\pi^-}$
data, I quote the average of $S_{\rho^+\rho^-}$ and the $\rho\pi$ Dalitz
analysis,
\beq\label{alpha}
\alpha = (103 \pm 11)^\circ .
\eeq
Including $\alpha$ extracted from $B\to\pi\pi$ would
make only a small difference at present, shifting $\alpha$ to
$(100^{+12}_{-10})^\circ$.
It is interesting to note that the direct determination of $\alpha$ in
Eq.~(\ref{alpha}) is already more precise than it is from the CKM fit, which
gives $\alpha = (98 \pm 16)^\circ$.

\subsection{$\gamma$ from $B^\pm\to DK^\pm$}

The idea is to measure the interference of $B^- \to D^0 K^-$ ($b\to c\bar u s$)
and $B^-\to \D0bar K^-$ ($b \to u \bar c s$) transitions, which can be studied
in final states accessible in both $D^0$ and $\D0bar$ decays~\cite{GL,GW}. In
principle, it is possible to extract the $B$ and $D$ decay amplitudes, the
relative strong phases, and the weak phase $\gamma$ from the data.

A practical complication is that the amplitude ratio 
\beq
r_B \equiv {A(B^-\to \D0bar K^-) \over A(B^-\to D^0 K^-)}\,
\eeq
is expected to be small. To make the two interfering amplitudes comparable in
size, the ADS method~\cite{ADS} was proposed to study final states where
Cabibbo-allowed and doubly Cabibbo-suppressed $D$ decays interfere.  While this
method is being pursued experimentally, some other recently proposed variants
may also be worth a closer look.  If $r_B$ is not much below $\sim$0.2, then
studying singly Cabibbo-suppressed $D$ decays may be advantageous, such as
$B^\pm \to K^\pm (K K^*)_D$~\cite{Grossman:2002aq}.  In three-body $B$ decays
the color suppression of one of the amplitudes can be avoided~\cite{APS}.

\begin{figure}
\centerline{\includegraphics[width=.45\textwidth]{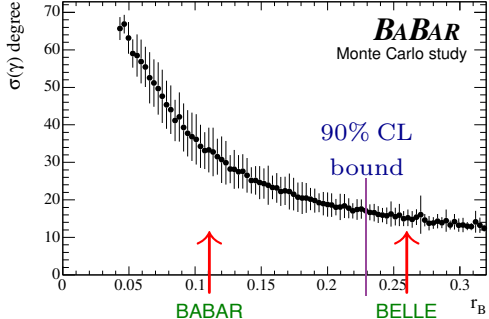}}
\caption{Monte Carlo study of the correlation of $r_B$ and the error of
$\gamma$~\cite{cavoto}.  The central values of the BABAR and BELLE measurements
are shown, together with an upper bound from the ADS analysis.}
\label{fig:gamma_rb}
\end{figure}

It was recently realized~\cite{bondar,Giri:2003ty} that both $D^0$ and $\D0bar$
have Cabibbo-allowed decays to certain 3-body final states, such as
$K_S\pi^+\pi^-$.  This analysis has only a two-fold discrete ambiguity, and one
can integrate over regions of the Dalitz plot, potentially enhancing the
sensitivity.  The best present determination of $\gamma$ comes from this
analysis.  BELLE obtained from 140\,\fbmo\ data~\cite{Poluektov:2004mf}
\beq
\gamma = 77^{+17}_{-19} \pm 13 \pm 11^\circ(model)\,,
\eeq
while BABAR found from 191\,\fbmo\ data~\cite{cavoto}
\beq
\gamma = 88 \pm 41 \pm 19 \pm 10^\circ(model)\,.
\eeq
The sizable difference in the errors in these measurements is due to the
large correlation between the error of $\gamma$ and the value of $r_B$, as shown
in Fig.~\ref{fig:gamma_rb}.  While BELLE found~\cite{bozek} $r_B = 0.26
^{+0.11}_{-0.15} \pm 0.03 \pm 0.04$, BABAR obtained $r_B < 0.18$ (90\% CL), with
the central values shown in Fig.~\ref{fig:gamma_rb}.  From the ADS analyses 90\%
CL upper bounds on $r_B$ were obtained, $r_B < 0.23$ at BABAR~\cite{cavoto} and
$r_B < 0.28$ at BELLE~\cite{bozek}.  These analyses are consistent with each
other at the $1-1.5\sigma$ level, but it will take more data to pin down $r_B$
and determine $\gamma$ more precisely.

\section{\boldmath Implications of the first $\alpha$ and $\gamma$ measurements}
\label{sec:NPmix}

\begin{figure}[t]
\centerline{\includegraphics[width=.45\textwidth]{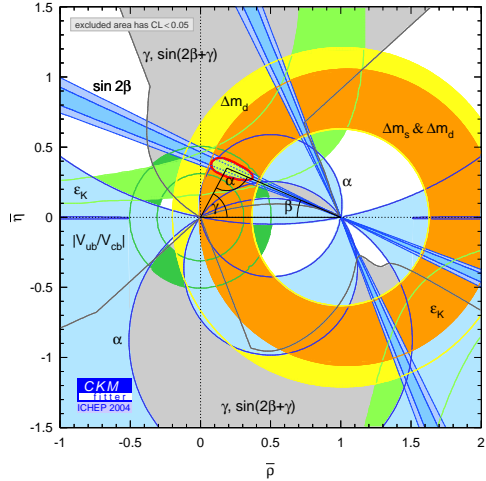}}
\caption{CKM fits including $\alpha$ and $\gamma$ measurements.}
\label{fig:smplotnew}
\end{figure}

Since the goal of the $B$ factories is to overconstrain the CKM matrix, one
should include in the CKM fit all measurements that are not limited by
theoretical uncertainties.  The result of such a fit is shown in
Fig.~\ref{fig:smplotnew}, which includes in addition to the inputs in
Fig.~\ref{fig:smplot} the following: (i) $\alpha$ from $B\to \rho\rho$ and from
the $\rho\pi$ Dalitz analysis, (ii) $\gamma$ from $B\to DK$ (with $D\to
K_S\pi^+\pi^-$), and (iii) $2\beta + \gamma$ from $B\to D^{(*)\pm}\pi^\mp$
measurements.

The best fit region in Fig.~\ref{fig:smplotnew} shrinks only slightly compared
to Fig.~\ref{fig:smplot}.  An interesting consequence of the new fit is a
noticeable reduction in the allowed range of $B_s-\Bbar_s$ mixing.  While the
standard CKM fit gives $\Delta m_s = \big(17.9 ^{+10.5} _{-1.7}\,
{}^{[+20.0]}_{[-2.8]}\big)\, {\rm ps}^{-1}$ at $1\sigma\ [2\sigma]$, the new fit
gives $\Delta m_s = \big(17.9 ^{+7.4} _{-1.4}\, {}^{[+13.3]}_{[-2.7]}\big)\,
{\rm ps}^{-1}$.

\subsection{New physics in $B^0-\B0bar$ mixing}

\begin{figure*}[t]
\centerline{\hspace{8pt}
\includegraphics[width=.47\textwidth,height=.44\textwidth]{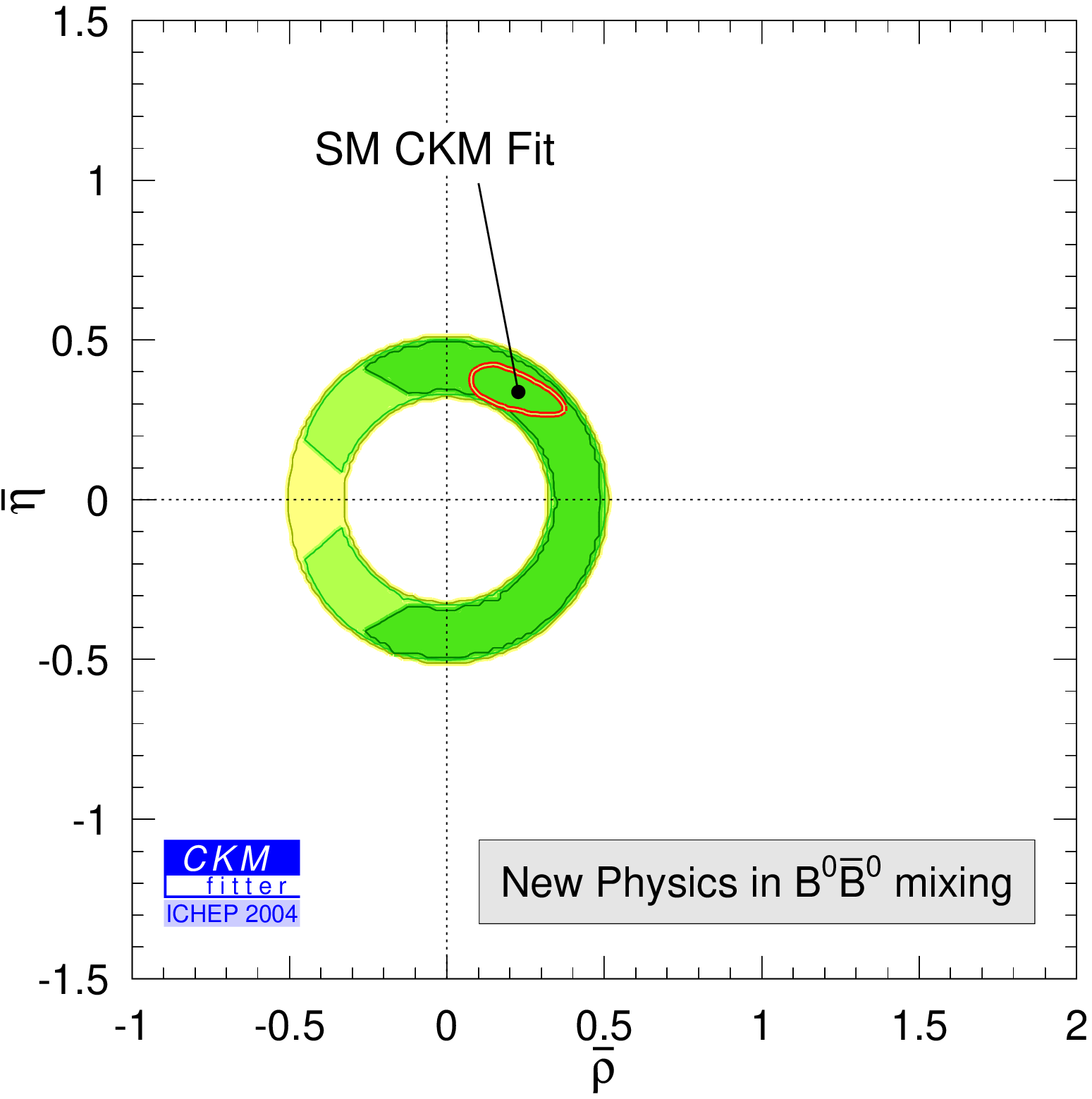}
\hspace{6pt} \includegraphics[width=.47\textwidth,height=.44\textwidth]{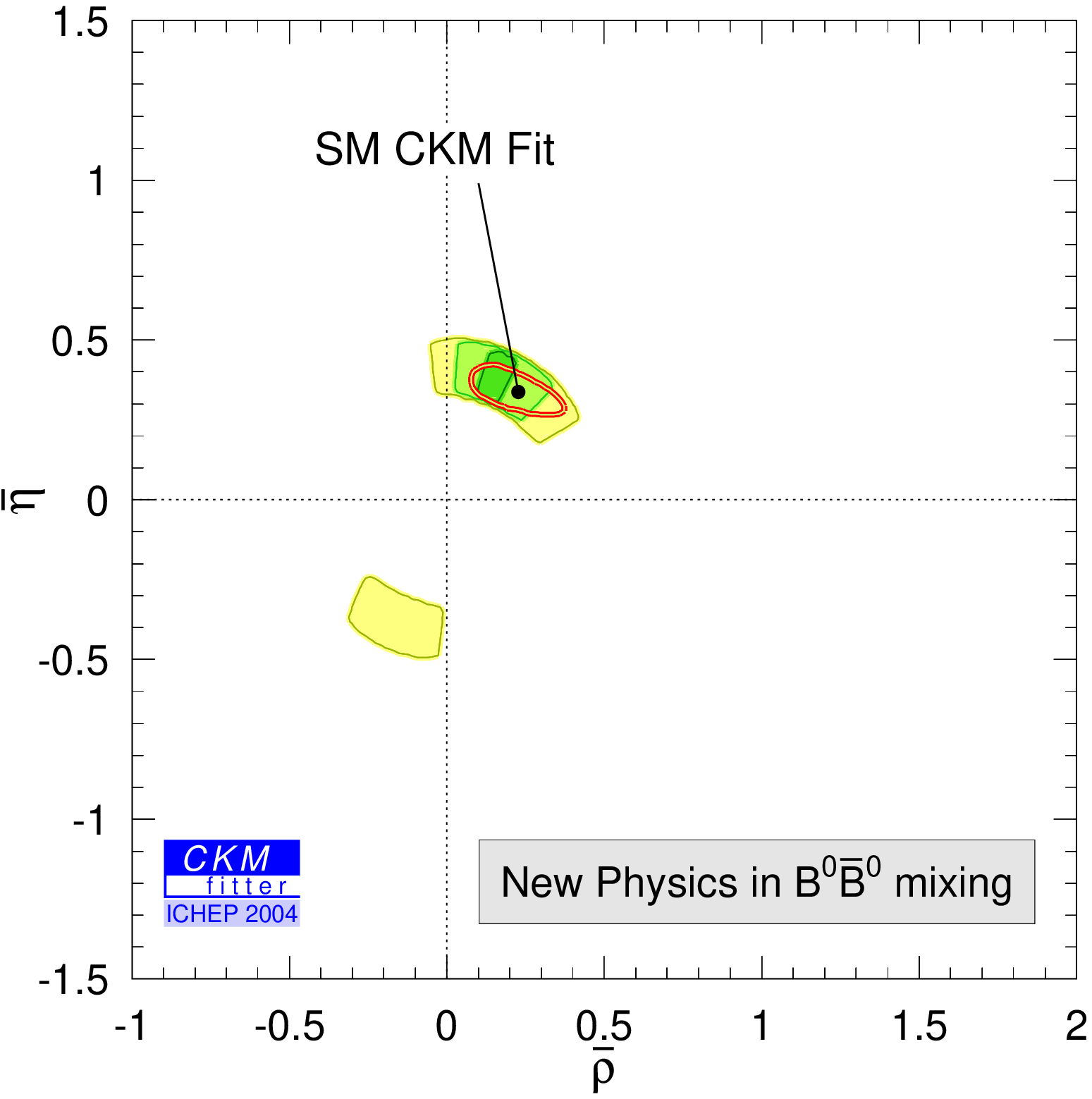}}
\vspace*{2pt}
\centerline{\includegraphics[width=.49\textwidth]{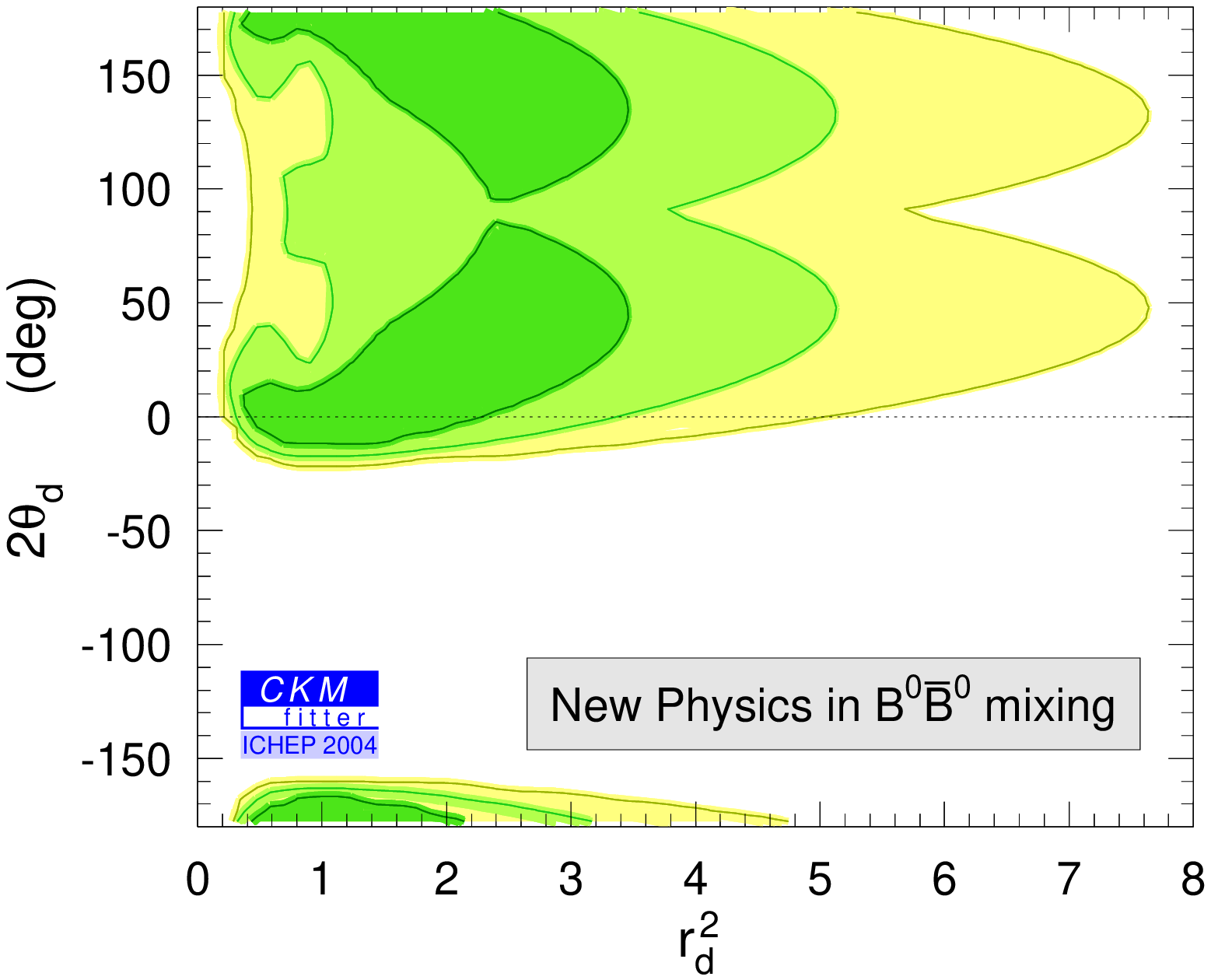}
\includegraphics[width=.49\textwidth]{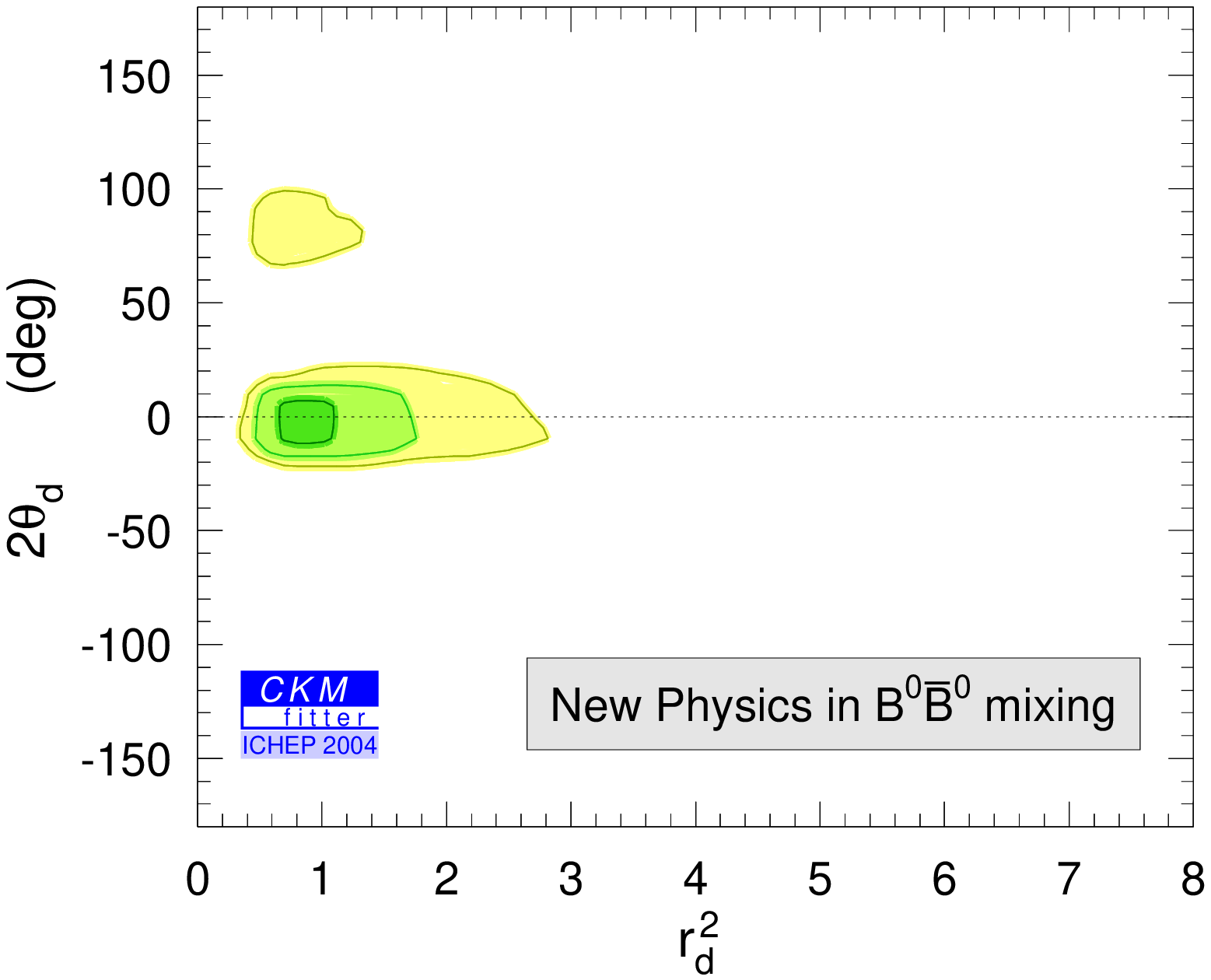}}
\caption{Allowed regions in the $\rho-\eta$ plane (top) and the
$r_d^2-2\theta_d$ plane (bottom) in the presence of new physics in $B-\Bbar$
mixing.  The left [right] plots are the allowed regions without [with] the new
constraints on $\alpha$, $\gamma$, $\cos2\beta$, and $2\beta+\gamma$.  The dark,
medium, and light shaded areas have CL $>$ 0.90, 0.32, and 0.05, respectively.}
\label{fig:npplot}
\end{figure*}

These new measurements give powerful constrain new physics.  In a large class of
models the dominant NP effect is to modify the $B^0-\B0bar$ mixing amplitude,
that can be parameterized~\cite{Grossman:1997dd} as $M_{12} = M_{12}^{\rm (SM)}
r_d^2\, e^{2i\theta_d}$.  Then, e.g., $\Delta m_B = r_d^2\, \Delta m_B^{\rm
(SM)}$, $S_{\psi K} = \sin(2\beta + 2\theta_d)$,  $S_{\rho^+\rho^-} =
\sin(2\alpha - 2\theta_d)$, while $|V_{ub}/V_{cb}|$ and $\gamma$ extracted from
$B\to DK$ are tree-level measurements which are unaffected.  Since $\theta_d$
drops out from $\alpha+\beta$, the measurements of $\alpha$, together with
$\beta$, are effectively equivalent in these models to NP-independent
measurements of $\gamma$ (up to discrete ambiguities).

Figure~\ref{fig:npplot} shows the fit results using only $|V_{ub}/V_{cb}|$,
$\Delta m_B$, $S_{\psi K}$, and $A_{\rm SL}$ as inputs (left) and also including
the measurements of $\alpha$, $\gamma$, $\cos2\beta$, and $2\beta+\gamma$
(right) in the $\rho-\eta$ plane (top) and the $r_d^2-2\theta_d$ plane
(bottom).  The new data determines $\rho$ and $\eta$ from (effectively)
tree-level $B$ decays, independent of mixing, and agrees with the other SM
constraints.  The allowed region in the $r_d^2-2\theta_d$\break parameter space
has shrunk immensely.  The somewhat disfavored ``non-SM region" around
$2\theta_d \sim 80^\circ$ (CL $< 20\%$) is due to  the $\eta<0$ region in the
top right plot and discrete ambiguities.  Thus, NP in $B^0-\B0bar$ mixing is
severely constrained now for the first time.

\section{\boldmath Theoretical developments}
\label{sec:scet}

$B$ physics is not only a great place to look for new physics, it also allows us
to study the interplay of weak and strong interactions in the SM at a level of
unprecedented detail.  There are many observables very sensitive to NP, and the
question is whether we can disentangle possible signals of NP from the hadronic
physics.  In the last few years there has been significant progress toward a
model independent theory of certain exclusive nonleptonic decays in the $m_B \gg
\lqcd$ limit.

While the theory of nonleptonic $B$ decays is most developed for heavy-to-heavy
decays of the type $B\to D^{(*)}\pi$~\cite{BBNS,BPSdpi,Mantry:2003uz} and
$\Lambda_b\to \Lambda_c \pi$ or $\Sigma_c\pi$~\cite{llsw}, here we concentrate
on charmless $B$ decays, as these are the most sensitive to new physics.  There
are several approaches.  The soft form factor and hard scattering contributions
are of the same order in the $1/m_b$ power counting.  Both Beneke {\it et
al.}~\cite{bbnslight} and Keum {\it et al.}~\cite{keumetal} make assumptions
about the $\alpha_s$ suppression of one or the other term.  An
SCET~\cite{Bauer:2000ew} analysis finds the two terms
comparable~\cite{Bauer:2004tj}, but predictive power is retained.

One of the most contentious issues is the role of charm
penguins~\cite{charmloops}, and whether strong phases are small. 
($A_{K^-\pi^+}$ in Eq.~(\ref{KpidirectCP}) tells us that some strong phases are
large.)  As far as I can tell, no suppression of the long distance part of charm
penguins has been proven.  In the absence of such a proof, we should view this
as a nonperturbative ${\cal O}(1)$ term that can give rise to many ``unexpected"
things, such as strong phases~\cite{Bauer:2004tj}.  (Note that whether one talks
about ``long distance charm loops", ``charming penguins", or ``$D\Dbar$
rescattering", it's all the same thing with different names.)

\subsection{Polarization in charmless $B\to VV$}
\label{sec:pol}

It has been argued~\cite{Kagan:2004uw} that the chiral structure of the SM and
the heavy quark limit imply that charmless $B$ decays to a pair of vector
mesons, such as $B\to \phi K^*$, $\rho\rho$, and $\rho K^*$ must have
longitudinal polarization fractions near unity, $f_L = 1 - {\cal O}(1/m_b^2)$. 
It is now well-established (see Table~\ref{tab:VVpol}) that in the penguin
dominated $\phi K^*$ modes $f_L \approx 0.5$.  We would like to know if this is
consistent with the SM.

\begin{table}[t]
\caption{Longitudinal polarization fractions in charmless $B\to VV$
decays~\cite{gritsan,zhang}.}
\label{tab:VVpol}
\centerline{\begin{tabular}{|c|cc|}
\hline
\raisebox{-6pt}[0pt][-6pt]{$B$ decay}  & 
  \multicolumn{2}{c|}{Longitudinal polarization}  \\
&  BELLE  &  BABAR  \\
\hline\hline
$\rho^- \rho^+$  &    &  $0.99 ^{+0.05}_{-0.04}$  \\
$\rho^0 \rho^+$  &  $0.95 \pm 0.11$  &  $0.97 ^{+0.06}_{-0.08}$  \\
$\omega \rho^+$  &    &  $0.88 ^{+0.12}_{-0.15}$  \\
\hline
$\rho^0 K^{*+}$  &    &  $0.96^{+0.06}_{-0.16}$  \\
$\rho^- K^{*0}$  &  $0.50 \pm 0.20$  &  $0.79 \pm 0.09$  \\
\hline
{$\phi K^{*0}$}  &  $0.52 \pm 0.08$  &  $0.52\pm 0.05$ \\
{$\phi K^{*+}$}  &  $0.49 \pm 0.14$  &  $0.46\pm 0.12$  \\
\hline
\end{tabular}}
\end{table}

Recently several explanations were proposed why the data may be consistent with
the SM~\cite{Bauer:2004tj,Kagan:2004uw,Colangelo:2004rd,Hou:2004vj}.  In SCET
the charm penguins, if they indeed have an unsuppressed long distance part, can
explain the data~\cite{Bauer:2004tj}.  The $D_s^{(*)} D^{(*)}$
rescattering~\cite{Colangelo:2004rd} can be viewed as a model calculation of
this effect.  It has also been argued that there are large ${\cal O}(1/m^2)$
effects from annihilation graphs~\cite{Kagan:2004uw}; however, if this is to
explain an ${\cal O}(1)$ effect in $f_L$ then the validity of the whole
expansion should be questioned.  Unfortunately it may be difficult to
experimentally distinguish between these two proposals, as they appear to enter
different rates in the same ratios.

While the $f_L(\phi K^*)$ data may be a result of a new physics contribution
(just like $A_{K^-\pi^+}$), we cannot rule out at present that it is simply due
to SM physics.

\subsection{$B\to K\pi$ branching ratios and $CP$ asymmetries}
\label{sec:Kpi}

$B\to K\pi$ decays are sensitive to the interference of $b\to s$ penguin and
$b\to u$ tree processes (and possible new physics).  The SM contributions that
interfere have different weak and possibly different strong phases, so the
challenge is if one can make sufficiently precise predictions to do sensitive
tests.

\begin{table}[t]
\caption{World average $CP$-averaged $B\to\pi K$ branching ratios, and $CP$
asymmetries.}
\label{tab:Kpi}
\centerline{\begin{tabular}{|l|ll|}  
\hline
\multicolumn{1}{|c|}{Decay mode}  &  \multicolumn{1}{c}{${\cal B}\ [10^{-6}]$}
  &  \multicolumn{1}{c|}{$A_{CP}$}
\\ \hline\hline
$\B0bar\to \pi^+K^-$  &  $18.2 \pm 0.8$  &  $-0.11 \pm 0.02$ \\
$B^-\to \pi^0K^-$  &  $12.1 \pm 0.8$  &  $+0.04 \pm 0.04$ \\
$B^-\to \pi^-\K0bar$  &  $24.1 \pm 1.3$  &  $-0.02 \pm 0.03$ \\
$\B0bar\to \pi^0\K0bar$  &  $11.5 \pm 1.0$  &  $+0.00 \pm 0.16$ \\ \hline
\end{tabular}}
\end{table}

The world average branching ratios and $CP$ asymmetries are shown in
Table~\ref{tab:Kpi}.  Besides the $5.7\sigma$ measurement of $A_{K^-\pi^+}$,
another interesting feature of the data is the $3.3\sigma$ difference,
$A_{K^-\pi^0} - A_{K^-\pi^+} = 0.15 \pm 0.04$.  This implies, assuming the SM,
that ``color-allowed" tree amplitudes do not dominate over ``color-suppressed"
trees (plus electroweak penguins).  While this may be a challenge for some
approaches, in SCET it is natural that color-allowed and -suppressed trees are
comparable in charmless $B$ decays~\cite{Bauer:2004tj}.

Concerning the branching ratios, I have been warned by several experimentalists
that their interpretation should be handled with care.\footnote{Until last
Summer, BABAR and BELLE used in these decays Monte Carlo simulations without
treatment of radiative corrections~\cite{Kpiwarning}.  This may underestimate
the rates for modes with light charged particles.  BABAR's new results include
corrections due to such effects, but ${\cal B}(\B0bar\to \pi^+K^-)$ has not been
updated.}  There are four ratios that have been extensively discussed in the
literature~\cite{Fleischer:1997um,Lipkin:1998ie,Grossman:1999av,%
Neubert:1998jq,Buras:2000gc,Yoshikawa:2003hb,Gronau:2003kj,Buras:2003yc},
\beqa\label{Kpiratios}
R_c &\equiv& 2\, \frac{{\cal B}(B^+\to \pi^0K^+) +
  {\cal B}(B^-\to \pi^0K^-)}{{\cal B}(B^+\to \pi^+ K^0) +
  {\cal B}(B^-\to \pi^- \K0bar)} \nn\\*
&=& 1.004 \pm 0.084\,, \nn\\
R_n &\equiv& \frac12\, \frac{{\cal B}(B^0\to \pi^- K^+) + 
  {\cal B}(\B0bar\to \pi^+ K^-)}{{\cal B}(B^0\to \pi^0K^0) + 
  {\cal B}(\B0bar\to \pi^0\K0bar)} \nn\\*
&=& 0.789 \pm 0.075\,, \nn\\
R &\equiv& \frac{\Gamma(B^0\to \pi^- K^+) +
  \Gamma(\B0bar\to \pi^+ K^-)}{\Gamma(B^+\to \pi^+ K^0) +
  \Gamma(B^-\to \pi^- \K0bar)} \nn\\*
&=& 0.820 \pm 0.056\,, \nn\\
R_L &\equiv& 2\, \frac{\bar\Gamma(B^-\to \pi^0 K^-) +
  \bar\Gamma(\B0bar\to \pi^0 \K0bar)}{\bar\Gamma(B^-\to \pi^- \K0bar) +
  \bar\Gamma(\B0bar\to \pi^+ K^-)} \nn\\*
&=& 1.123 \pm 0.070\,,
\eeqa
where $\Gamma \equiv {\cal B}/\tau$, and $\bar\Gamma$ in the last equation
denotes the $CP$-averaged widths.  These ratios are interesting, as their
deviations from unity are sensitive to different corrections to the dominant
penguin amplitudes.

The pattern of these ratios is quite different from what it was before ICHEP:
$R_c$ and $R_L$ became significantly closer to unity, while $R$'s deviation from
unity increased.  This seems to disfavor new physics
explanations~\cite{Buras:2003yc,Grossman:1999av}, according to which NP
primarily modifies electroweak penguin contributions.  This is because
electroweak penguins are color allowed in the modes involving $\pi^0$'s, that
enter $R_c$, while they are color suppressed in the other ones, such as those in
$R$.  

Since $R$ is significantly below unity, at present the Fleischer-Mannel
bound~\cite{Fleischer:1997um} is interesting again, giving $\gamma < 75^\circ$
(95\% CL).

It will be fascinating to understand the theory in sufficient detail to sort out
what the data is telling us, and also to see where the measurements will settle.

\section{\boldmath Outlook}
\label{sec:fin}

\begin{table*}[t]
\caption{Some interesting measurement that are far from being theory limited. 
The errors for the $CP$ asymmetries in the first box refer to the angles in
parenthesis, assuming typical values for other parameters.}
\label{tab:future}
\centerline{
\begin{tabular}{|l|c|c|}
\hline
Measurement (in SM)  &  Theoretical limit  &  Present error  \\ 
\hline\hline 
$B\to \psi K_S$ \ ($\beta$)  &  $\sim 0.2^\circ$  &  $1.6^\circ$ \\
$B\to \phi K_S,\ \eta^{(\prime)}K_S$, ... ($\beta$)  
  &  $\sim 2^\circ$  &  $\sim 10^\circ$\\
$B\to \pi\pi,\ \rho\rho,\ \rho\pi$ \ ($\alpha$)  
  &  $\sim 1^\circ$  &  $\sim15^\circ$\\
$B\to DK$ \ ($\gamma$)  &  $\ll 1^\circ$  &  $\sim 25^\circ$  \\
$B_s\to \psi\phi$ \ ($\beta_s$)  &  $\sim 0.2^\circ$  &  --- \\ 
$B_s\to D_sK$ \ ($\gamma-2\beta_s$)  &  $\ll 1^\circ$  &  --- \\
\hline
$|V_{cb}|$  &  $\sim 1\%$  &  $\sim3\%$ \\
$|V_{ub}|$  &  $\sim 5\%$  &  $\sim15\%$ \\
$B\to X_s \gamma$  &  $\sim 5\%$  &  $\sim10\%$ \\
$B\to X_s \ell^+\ell^-$  &  $\sim 5\%$  &  $\sim20\%$ \\
$B\to X_s\nu\bar\nu, K^{(*)} \nu\bar\nu$  &  $\sim 5\%$ &  ---  \\
\hline
$K^+\to \pi^+\nu\bar\nu$  &  $\sim 5\%$  &  $\sim70\%$ \\
$K_L\to \pi^0\nu\bar\nu$  &  $< 1\%$  &  --- \\
\hline
\end{tabular}}
\end{table*}

Having seen these impressive measurements, one should ask where we go from here
in flavor physics?  Whether we see in the next few years stronger signals of
flavor physics beyond the SM will certainly be decisive~\cite{Ligeti:2002wt}. 
The existing measurements could have shown deviations from the SM, and if there
are new particles at the TeV scale, new flavor physics could show up ``any
time".  In fact, we do not know whether we are seeing hints or just statistical
fluctuations in the $S_{b\to s}$ data.

For BABAR and BELLE, reducing the error of $S_{\psi K}$ to the few percent level
has been a well-defined target.  The data sets have roughly doubled each year
for the past several years, and will reach $500-1000\,\fbmo$ each in a few
years, possibly allowing for unambiguous observation of NP if the central values
do not change too much.  If NP is seen in flavor physics then we will certainly
want to study it in as many different processes as possible.  If NP is not seen
in flavor physics, then it is interesting to achieve what is theoretically
possible, thereby testing the SM at a much more precise level.  Even in the
latter case, flavor physics will give powerful constraints on model building in
the LHC era.  

The present status and (my estimates of) the theoretical limitations of some of
the theoretically cleanest measurements are summarized in
Table~\ref{tab:future}.  It shows that the sensitivity to NP is not limited by
hadronic physics in many measurements for a long time to come.  One cannot
overemphasize that the program as a whole is a lot more interesting than any
single measurement, since it is the multitude of ``overconstraining"
measurements and their correlations that are likely to carry the most
interesting information.

\section{Conclusions}
\label{sec:conc}

The large number of impressive new results speak for themselves, so it is easy
to summarize the main lessons we have learned:

\begin{itemize}\itemsep -2pt

\item $\sin2\beta = 0.726 \pm 0.037$\\
implies that the overall consistency of the SM is very good, and the KM phase is
probably the dominant source of CPV in flavor changing processes;

\item $S_{\psi K} - \langle -\eta_f S_{f(b\to s)} \rangle = 0.30\pm0.08\
(3.5\sigma)$ and $S_{\psi K} - S_{\eta'K_S} = 0.31 \pm 0.12\ (2.6\sigma)$\\
imply that we may be observing hints of NP in $b\to s$ transitions, since the
present central values with $5\sigma$ would be quite convincing;

\item $A_{K^- \pi^+} = -0.11 \pm 0.02\ (5.7\sigma)$\\
implies that ``$B$-superweak" models are excluded and that there are large
strong phases in some charmless $B$ decays;

\item First measurements of $\alpha$ and $\gamma$ \\
imply that the direct measurement of $\alpha$ is already more precise than the
indirect CKM fit, and finally we have severe constraints on NP in $B-\Bbar$
mixing.

\end{itemize}

\section*{Acknowledgments}

I am grateful to Andreas H\"ocker and Heiko Lacker for working out the CKM fits,
Tom Browder and Jeff Richman for help with the experimental data, and all of
them, Gilad Perez, and especially Yossi Nir for many interesting discussions.
I thank the organizers for the invitation to a very enjoyable conference.
This work was supported in part by the Director, Office of Science,
Office of High Energy and Nuclear Physics, Division of High Energy Physics, of
the  U.S.\ Department of Energy under Contract DE-AC03-76SF00098 and by a DOE
Outstanding Junior Investigator award.

\end{document}